\documentclass[journal=jacsat,manuscript=article]{achemso}
\usepackage[T1]{fontenc}
\usepackage{amsmath}
\usepackage{xcolor}

\author{Jia-Liang Xie}
\affiliation[First Address]
{State Key Laboratory of Semiconductor Physics and Chip Technologies, Institute of Semiconductors, Chinese Academy of Sciences, Beijing 100083, China}
\alsoaffiliation[Second Address]
{Center of Materials Science and Optoelectronics Engineering, University of Chinese Academy of Sciences, Beijing, 100049, China}
\author{Ting-Ting Wang}
\affiliation[First Address]
{Phonon Engineering Research Center of Jiangsu Province, Ministry of Education Key Laboratory of NSLSCS, Center for Quantum Transport and Thermal Energy Science, Institute of Physics Frontiers and Interdisciplinary Sciences, School of Physics and Technology\\ Nanjing Normal University\\ Nanjing 210023, China}
\author{Chen-Kai Liu}
\affiliation[First Address]
{State Key Laboratory of Semiconductor Physics and Chip Technologies, Institute of Semiconductors, Chinese Academy of Sciences, Beijing 100083, China}
\alsoaffiliation[Second Address]
{Center of Materials Science and Optoelectronics Engineering, University of Chinese Academy of Sciences, Beijing, 100049, China}
\author{Rui Mei}
\affiliation[First Address]
{State Key Laboratory of Semiconductor Physics and Chip Technologies, Institute of Semiconductors, Chinese Academy of Sciences, Beijing 100083, China}
\alsoaffiliation[Second Address]
{Center of Materials Science and Optoelectronics Engineering, University of Chinese Academy of Sciences, Beijing, 100049, China}
\author{Li-Fa Zhang}
\affiliation[First Address]
{Phonon Engineering Research Center of Jiangsu Province, Ministry of Education Key Laboratory of NSLSCS, Center for Quantum Transport and Thermal Energy Science, Institute of Physics Frontiers and Interdisciplinary Sciences, School of Physics and Technology\\ Nanjing Normal University\\ Nanjing 210023, China}
\author{Miao-Ling Lin}
\affiliation[First Address]
{State Key Laboratory of Semiconductor Physics and Chip Technologies, Institute of Semiconductors, Chinese Academy of Sciences, Beijing 100083, China}
\alsoaffiliation[Second Address]
{Center of Materials Science and Optoelectronics Engineering, University of Chinese Academy of Sciences, Beijing, 100049, China}
\email{linmiaoling@semi.ac.cn}
\author{Ping-Heng Tan}
\affiliation[First Address]
{State Key Laboratory of Semiconductor Physics and Chip Technologies, Institute of Semiconductors, Chinese Academy of Sciences, Beijing 100083, China}
\alsoaffiliation[Second Address]
{Center of Materials Science and Optoelectronics Engineering, University of Chinese Academy of Sciences, Beijing, 100049, China}
\email{phtan@semi.ac.cn}

\title[Running title]
{Birefringence-Driven Anisotropic $\alpha$-MoO$_3$ Optical Cavities}

\begin{document}

\begin{abstract}
Many anisotropic layered materials, despite their strong in-plane birefringence, exhibit substantial visible absorption, which severely restricts cavity lengths and hinders the observation of purely birefringence-governed optical phenomena. Here, we realize a birefringence-driven anisotropic optical cavity using $\alpha$-MoO$_3$ flakes, capitalizing on their ultralow optical loss and pronounced in-plane birefringence. Using angle-resolved polarized Raman (ARPR) spectroscopy, we observe a mode-sensitive enhancement of anisotropy, dependent on both flake thickness and Raman shift. A unified model that incorporates the intrinsic Raman tensor, birefringence, and chromatic dispersion accurately reproduces the experimental data, elucidating how cavity resonances at both excitation and scattered wavelengths interact. Within this framework, the intrinsic phonon anisotropy is quantified, providing invaluable insights for accurately predicting ARPR responses and identifying crystallographic orientation. This work provides fundamental insights into birefringence-governed cavities and opens avenues for high-performance birefringent optics and cavity-enhanced anisotropic phenomena.

\end{abstract}

\section{Introduction}\label{sec1}

Anisotropic layered materials (ALMs) exhibit exceptional in-plane optical anisotropy\cite{Qiao-2014-NC,Zhang-np-2022,Ermolaev-nc-2021}, providing unprecedented control over light-matter interactions\cite{Biswas-sci-2021,Zhang-np-2022,huang-sci-2024} through their anisotropic complex refractive index, characterized by birefringence and linear dichroism\cite{Qiao-2014-NC,Zhang-np-2022,Song-nat-2022,Guo-np-2024}. This fundamental property not only enables novel photonic and optoelectronic devices\cite{Biswas-sci-2021,Yuan-np-2021,Chen-nc-2024} but also offers unique spectroscopic fingerprints through Raman scattering\cite{Ribeiro-2015-ACSNano,Li-2019-AM,Mao-2019-JACS,Zhang-2022-NC,Xie-2025-AM}. The integration of ALMs into optical cavities, formed naturally on dielectric substrates\cite{Biswas-sci-2021,Zhang-np-2022,Dereshgi-2020-NC}, creates hybrid systems where the material's intrinsic anisotropy is imprinted onto cavity modes\cite{Zhang-np-2022}, establishing a powerful platform for anisotropic  modulation on photonic processes in low-symmetry systems. This cavity-ALM synergy represents a significant advancement beyond conventional isotropic material systems\cite{Lin-PRL-2025}, offering new dimensions for controlling light-matter interactions.

A fundamental paradox confronts many prominent narrow-bandgap ALMs: despite their exceptionally strong birefringence and linear dichroism, yet suffering from significant optical dissipation in the visible spectrum. This limitation, observed in materials such as black phosphorus\cite{Qiao-2014-NC,Lin-2020-SciBull,Xie-2025-AM}, transition metal dichalcogenides (TMDs)\cite{McCreary-2017-NL,Gu-2020-AM,Ermolaev-nc-2021}, and quasi-1D materials\cite{Niu-np-2018,Wu-2019-AdvM}, severely curtails the effective cavity length and directly restricts achievable modulation in anisotropic Raman scattering. To overcome this absorption bottleneck, an ALM with ultralow optical absorption and thus negligible linear dichroism is required to unlock a distinct birefringence-driven cavity effect, where the cavity response is dominated purely by birefringence$-$an effect critical for designing cavity-enhanced low-loss birefringent nanophotonics and deciphering fundamental mechanisms anisotropic light-matter interactions. Accordingly, $\alpha$-phase molybdenum trioxide ($\alpha$-MoO$_3$) emerges as a promising ALM\cite{Ma-2018-Nature,Dereshgi-2020-NC}, exhibiting exceptional phonon diversity\cite{Ma-2018-Nature,Chen-2020-NM,Dereshgi-2020-NC}, and critically, drastically low optical absorption\cite{Balendhran-2013-AM} with ultrabroadband birefringence across visible and infrared regions\cite{Shen-2023-2DM,Dereshgi-2023-AOM,Guo-2025-ASS}. Leveraging this extreme in-plane birefringence, prior studies have demonstrated highly anisotropic polaritons \cite{Ma-2018-Nature,Chen-2020-NM} and subwavelength polarization/phase optical applications\cite{Dereshgi-2020-NC}, underscoring $\alpha$-MoO$_3$ as an established materials platform for controlling low-loss anisotropic light-matter interactions. Furthermore, the wavelength-dependent optical anisotropy of $\alpha$-MoO$_3$ and its wavelength dependence have been quantitatively revealed\cite{Andres-2021-APL,Puebla-npj2DM}, providing a solid foundation for the cavity engineering herein. This unique property portfolio establishes $\alpha$-MoO$_3$ as an unparalleled platform to overcome cavity-length limitations and enable exploration of purely birefringence-driven cavity effects in anisotropic light-matter interactions such as Raman scattering.

In this work, we report a birefringence-driven anisotropic $\alpha$-MoO$_3$ optical cavity, directly observed through angle-resolved polarized Raman (ARPR) spectroscopy. This cavity effect produces a Raman mode-sensitive enhancement of anisotropy that depends critically on both the $\alpha$-MoO$_3$ thickness ($d_{\rm MoO_3}$) and the wavelength of the scattered light. Remarkably, this modulation of the ARPR intensity remains pronounced even in micron-thick crystals, starkly contrasting with the behaviour observed in strongly absorptive anisotropic materials. The entire suite of intriguing anisotropic Raman responses is well captured by a unified model incorporating the intrinsic Raman tensor ($\textit{\textbf{R}}^{\rm int}$), the photon wavelength of Raman scattering, and its chromatic dispersion in the birefringent $\alpha$-MoO$_3$ crystal. This model not only enables quantitative prediction of ARPR intensities in ultra-thick $\alpha$-MoO$_3$ flakes but also permits precise, unambiguous determination of its crystallographic orientation using the highly anisotropic $A_{\rm g}^2$ mode. Based on the established birefringence picture in $\alpha$-MoO$_3$, this work reveals a refined framework for the multi-photon cavity behaviors of Raman scattering, and it further demonstrates the thickness-tunable birefringence-driven cavity as an independent degree of freedom for modulating anisotropic light-matter interactions and engineering polarization functionality in birefringent ALM.

\section{Results and discussion}\label{sec2}
\subsection{Mode-sensitive anisotropic enhancement of ARPR intensity in $\alpha$-MoO$_3$}
\begin{figure*}[t]
  {\includegraphics[width=\linewidth,clip]{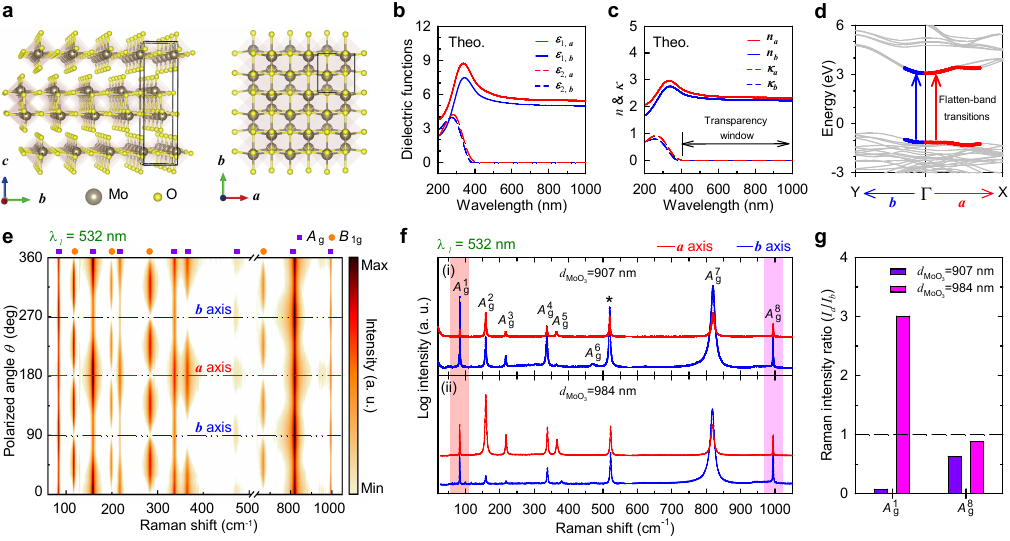}}
  \caption{a) Crystal structure of $\alpha$-MoO$_3$ in perspective view and along the $c$-axis, highlighting the primitive cell (black box) and MoO$_6$ octahedra (brown). b) Calculated in-plane dielectric functions ($\epsilon_1+{\rm i}\epsilon_2$) and c) complex refractive indices ($\tilde{n} = n+{\rm i}\kappa$) along the $a$-axis (red) and $b$-axis (blue). d) Electronic band structure along $\Gamma\text{-}Y$ and $\Gamma\text{-}X$. e) ARPR mapping of bulk-like $\alpha$-MoO$_3$ with labeled Raman modes. f) Polarized Raman spectra under $\textbf{\textit{e}}_l \parallel \textbf{\textit{e}}_s \parallel a$ and $\textbf{\textit{e}}_l \parallel \textbf{\textit{e}}_s \parallel b$ configurations for 907 nm and 984 nm flakes, respectively. $*$ denotes the Si Raman signal from the substrate. g) Corresponding experimental $I_a/I_b$ for the $A^1_{\rm g}$ and $A^8_{\rm g}$ modes. $\lambda_l = 532\ \text{nm}$.}\label{Fig1}
\end{figure*}

$\alpha$-MoO$_3$ crystallizes as a centrosymmetric biaxial ALM with a layered orthorhombic structure ($\textbf{Figure}$ \ref{Fig1}a)\cite{KIHLBORG-1963-AFK}. Each layer consists of MoO$_6$ octahedra that forms chain arrangements along the $a$-axis via corner-sharing and along the $b$-axis via edge-sharing\cite{KIHLBORG-1963-AFK}. The dielectric tensor in the crystallographic frame is diagonal, with principal components $\tilde{\epsilon}_{a(b,c)} = \epsilon_{1,a(b,c)}+{\rm i}\epsilon_{2,a(b,c)}$ along the $a$-, $b$-, and $c$-axes confirming its strong optical anisotropy (Figure \ref{Fig1}b,c)\cite{Lajaunie-2013-PRB}. Density functional theory (DFT) calculations reveal a giant indirect bandgap of $\sim$3.25 eV (Supplementary Figure S1) and pronounced anisotropic dispersion between the $\Gamma$-X and $\Gamma$-Y directions (Figure \ref{Fig1}d). With minimal absorption in the visible regime, the optical anisotropy primarily manifests as birefringence ($n_a\neq n_b$) dominated by the difference between $n_a$ and $n_b$ (Figure \ref{Fig1}c).

The visible-range transparency of $\alpha$-MoO$_3$ offers an ideal platform to investigate ARPR spectra modulated purely by birefringence. First-principles phonon dispersion calculations confirm 24 Raman-active modes in bulk $\alpha$-MoO$_3$ (Supplementary Figure S1c), with irreducible representations $\Gamma_{\rm Raman}$ = $8A_{\rm g} + 4B_{\rm 1g} + 8B_{\rm 2g} + 4B_{\rm 3g}$, spanning frequencies up to $\sim$ 1000 cm$^{-1}$.\cite{Gong-2022-AOM} We measured ARPR spectra in a parallel polarization configuration ($\textbf{\textit{e}}_l = \textbf{\textit{e}}_s = (\cos\theta, \sin\theta, 0)^{\rm T}$), where $\textbf{\textit{e}}_l$ and $\textbf{\textit{e}}_s$ denote the experimentally configured polarization states of incident laser ($l$) and scattered ($s$) light external to the sample, respectively. The angle $\theta$ between $\textbf{\textit{e}}_l$ (and $\textbf{\textit{e}}_s$) and the $a$-axis was controlled using a half-wave plate, with $\theta=0^\circ$ corresponding to $\textbf{\textit{e}}_l\parallel\textbf{\textit{e}}_s\parallel a$. Under 532 nm excitation (Figure \ref{Fig1}e), ARPR spectroscopy identified 12 phonon modes related to in-plane polarizability changes, corresponding to 8 $A_{\rm g}$ (purple) and 4 $B_{\rm 1g}$ (orange) modes\cite{Gong-2022-AOM}. The high-frequency modes primarily originate from stretching vibrations of the oxygen atoms in the lattice. By establishing the lab coordinate system ($xyz$) with $x\parallel a$, $y\parallel b$ and leaving $z$ perpendicular to the basal plane of $\alpha$-MoO$_3$, the Raman tensors for these modes are as follows,\cite{Loudon-1964}

\begin{equation}
  \begin{aligned}
  \quad\textit{\textbf{R}}(A_{\rm g})&=\begin{pmatrix}
                       |a|{\rm e}^{{\rm i}\phi_a} & 0 & 0 \\
                       0 & |b|{\rm e}^{{\rm i}\phi_b} & 0 \\
                       0 & 0 & |c|{\rm e}^{{\rm i}\phi_c} \\
                      \end{pmatrix},\\
  \quad\textit{\textbf{R}}(B_{\rm 1g})&=\begin{pmatrix}
                       0 & |d|{\rm e}^{{\rm i}\phi_d} & 0 \\
                       |d|{\rm e}^{{\rm i}\phi_d} & 0 & 0 \\
                       0 & 0 & 0 \\
                     \end{pmatrix}.
  \label{tensors}
  \end{aligned}
  \end{equation}

\noindent The measured ARPR intensity follows the Raman selection rule $I(\theta)\propto |\textit{\textbf{e}}_{s}^\dagger \cdot \textit{\textbf{R}}^{\rm eff} \cdot  \textit{\textbf{e}}_{l}|^2$\cite{Xie-2025-AM}, where $\textit{\textbf{R}}^{\rm eff}$ denotes the effective Raman tensor (e.g., $|a^{\rm eff}|{\rm e}^{{\rm i}\phi^{\rm eff}_a}$) that incorporates both the intrinsic phonon anisotropy and intricate optical modulation\cite{Xie-2025-AM}. This yields $I(\theta)\propto |a^{\rm eff}|^2 \cos^4\theta+|b^{\rm eff}|^2\sin^4\theta+2|a^{\rm eff}||b^{\rm eff}|\cos^2\theta\sin^2\theta\cos(\phi^{\rm eff}_a-\phi^{\rm eff}_b)$ and $|d^{\rm eff}|^2\sin^22\theta$ for the $A_{\rm g}$ and $B_{\rm 1g}$ modes, respectively. Thus, the ARPR intensities of $A_{\rm g}$ and $B_{\rm 1g}$ modes are expected to exhibit two-fold and four-fold rotational symmetries\cite{Gong-2022-AOM,Zou-2023-Small}, respectively, over the angular range of 0$^\circ$ to 360$^\circ$. Accordingly, the eight $A_{\rm g}$ modes are identified at 84, 158, 217, 337, 365, 471, 819, and 995 cm$^{-1}$, consistent with theoretical calculations in Supplementary Table S1. For clarity, these modes are labeled sequentially from $A^1_{\rm g}$ to $A^8_{\rm g}$ in order of increasing phonon frequency.

The optical response of in-plane anisotropic $\alpha$-MoO$_3$ is strongly correlated with $d_{\rm MoO_3}$, similar to the case of black phosphorus in our previous study\cite{Xie-2025-AM}. Polarized Raman spectroscopy ($\textbf{\textit{e}}_l\parallel\textbf{\textit{e}}_s\parallel a$ or $b$) of $\alpha$-MoO$_3$ flakes with $d_{\rm MoO_3}$ = 907 nm and 984 nm (Supplementary Figure S2), supported on 90 nm-SiO$_2$/Si substrates and excited with a 532 nm laser ($\lambda_l$ = 532 nm) (Figure \ref{Fig1}f), reveals a pronounced dependence of both the Raman intensity ($I$) and the anisotropy ratio ($I_a/I_b$) on $d_{\rm MoO_3}$ for each $A_{\rm g}$ mode, where $I_a$ and $I_b$ are the Raman intensity of a specific $A_{\rm g}$ mode when $\textbf{\textit{e}}_l\parallel\textbf{\textit{e}}_s\parallel a$ and $\textbf{\textit{e}}_l\parallel\textbf{\textit{e}}_s\parallel b$, respectively. This behavior can be intuitively understood in terms of the birefringence of $\alpha$-MoO$_3$ and the multilayer interference within the $\alpha$-MoO$_3$/SiO$_2$/Si heterostructure, as comprehensively analyzed in our earlier work\cite{Xie-2025-AM}. The $I_a/I_b$ ratios of different $A_{\rm g}$ modes are generally expected to exhibit similar oscillatory behavior with increasing $d_{\rm MoO_3}$ due to a common interference mechanism\cite{Xie-2025-AM}. However, a detailed comparison between different modes reveals a remarkable deviation and critically depends on the Raman shift. For example, the $A^1_{\rm g}$ and $A^2_{\rm g}$ mode follow a similar thickness evolution of $I_a/I_b$, while higher-frequency $A^7_{\rm g}$ and $A^8_{\rm g}$ modes behaves quite differently. To explicitly illustrate this dependence, we then select the $A^1_{\rm g}$ (84 cm$^{-1}$) and $A^8_{\rm g}$ (995 cm$^{-1}$) modes for detailed analysis, as they exhibit the largest difference in Stokes shift among all observed $A_{\rm g}$ modes, thereby providing the most pronounced contrast for examining how Raman shift influences cavity-mediated Raman intensity anisotropy. As shown in Figure \ref{Fig1}g, the $I_a/I_b$ ratio of the $A_{\rm g}^1$ mode increases by more than 40-fold when $d_{\rm MoO_3}$ increases from 907 nm (purple bar) to 984 nm (pink bar), resulting in a complete reversal of anisotropy. In contrast, the $I_a/I_b$ ratio of the $A_{\rm g}^8$ mode remains nearly constant, attributable to comparable intensity variations along both the $a$- and $b$-axes (Figure \ref{Fig1}g). This mode-sensitive anisotropic enhancement of Raman intensity highlights the complex role of birefringence on polarized Raman responses and underscores the necessity for a deeper understanding of ARPR spectroscopy in $\alpha$-MoO$_3$.

\subsection{Birefringence-driven anisotropic optical cavity in $\alpha$-MoO$_3$ flakes}

\begin{figure*}[!t]
  {\includegraphics[width=\linewidth,clip]{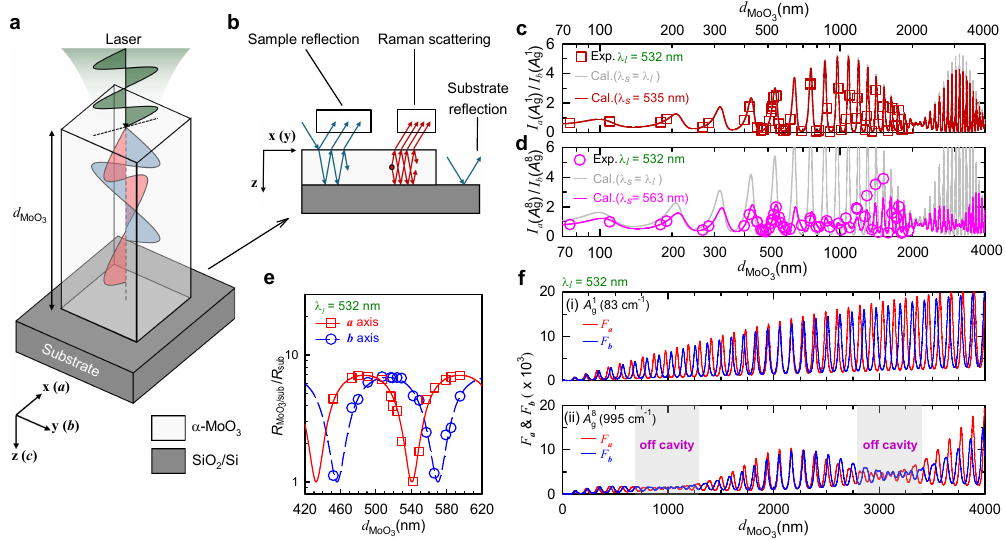}}
  \caption{a) Schematic of birefringent light propagation in $\alpha$-MoO$_3$, showing the initial polarization state (green) and its decomposed in-plane components along the $a$-axis (red) and $b$-axis (blue). b) Propagation paths of incident laser (blue) and Raman signal (red) within the $\alpha$-MoO$_3$ cavity, indicating reflection and scattering processes. Experimental (symbols) and calculated (curves) $I_a/I_b$ for the c) $A_{\rm g}^1$ and d) $A_{\rm g}^8$ modes. Gray curves assume $\lambda_{s}=\lambda_{l}$. e) Normalized reflectance ($R_{\rm MoO_3/Sub}$/$R_{\rm Sub}$) along the $a$- and $b$-axes versus $d_{\rm MoO_3}$ at $\lambda_{l}$ = 532 nm. Symbols: experimental data; curves: theoretical fits. f) Calculated $F_a$ and $F_b$ for the (i) $A_{\rm g}^1$ and (ii) $A_{\rm g}^8$ modes. Shaded regions indicate the off-cavity regime. }\label{Fig2}
\end{figure*}

$\textbf{Figure}$ \ref{Fig2}a shows a schematic illustration of birefringence ($\Delta n = |n_a - n_b|$) of the incident light within $\alpha$-MoO$_3$ flakes on a substrate. A similar behavior applies to the Raman-scattered light. Due to the difference between $n_a$ and $n_b$, the electric field components of the light propagate with distinct phase velocities along the two orthogonal polarization directions ($x(a)$ and $y(b)$). This results in a depth ($z$)-dependent phase retardation between the two components, altering the polarization state of the incident light inside the $\alpha$-MoO$_3$ flakes\cite{Xie-2025-AM,Lin-2020-SciBull,Antonacci-2024-NC}. Moreover, the complex refractive index ($\tilde{n}$) mismatch at the interfaces of the air/$\alpha$-MoO$_3$/SiO$_2$/Si multilayer causes the $\alpha$-MoO$_3$ flakes to act as a Fabry-P${\rm \acute{e}}$rot cavity\cite{Xie-2025-AM,Lin-PRL-2025} (Figure \ref{Fig2}b), significantly enhancing or suppressing the electric fields of the incident light and the internally generated Raman-scattered light through optical interference.

For an analysis of the cavity effects on the Raman intensity, we systematically investigated the $I_a/I_b$ ratios of the $A_{\rm g}^1$ and $A_{\rm g}^8$ modes as a function of $d_{\rm MoO_3}$ (Figure \ref{Fig2}c,d). For both Raman modes, the $I_a/I_b$ ratios exhibit pronounced interference oscillations with $d_{\rm MoO_3}$, demonstrating modulation by the Fabry-P${\rm \acute{e}}$rot cavity\cite{Xie-2025-AM,Lin-PRL-2025} (Figure \ref{Fig2}b) formed by the $\alpha$-MoO$_3$ flakes on the SiO$_2$/Si substrate. To quantify the cavity modulation on the Raman-scattered light excited by the incident laser at $\lambda_{l}$ = 532 nm, we determined its in-plane $n_a$, $n_b$ by the polarized laser reflectance\cite{Xie-2025-AM} ($R_{{\rm MoO_3/Sub},p}, p = a, b$) of $\alpha$-MoO$_3$/90 nm-SiO$_2$/Si with various $d_{\rm MoO_3}$ (see experimental details in Supplementary Section 1). The measured reflectance $R_{{\rm MoO_3/Sub},p}$ ($p = a, b$) relative to that ($R_{\rm Sub}$) of the bare 90 nm-SiO$_2$/Si substrate as a function of $d_{\rm MoO_3}$  (Figure \ref{Fig2}e) can be well fitted by the derived formula including $n_a$ and $n_b$ (Supplementary Section 2) based on the transfer matrix method (TMM), yielding $n_a$ = 2.46 and $n_b$ = 2.33, as listed in Supplementary Table S2. The extracted birefringence $\sim$0.13 is consistent with literature\cite{Andres-2021-APL,Puebla-npj2DM,Shen-2023-2DM}. The absence of imaginary parts in complex refractive indexes along $a$ and $b$ axes further confirms the optical cavity effect purely driven by birefringence for incident and Raman-scattered light inside $\alpha$-MoO$_3$.

The Raman signal scattered into air from phonon modes inside $\alpha$-MoO$_3$ flakes is determined by the electric field components of the excitation and Raman-scattered light at each depth $z$, both of which undergo modulation by Fabry-P${\rm \acute{e}}$rot cavity\cite{Xie-2025-AM,Lin-PRL-2025} driven by birefringence. Prior work on anisotropic ALM flakes established a comprehensively quantitative model\cite{Xie-2025-AM} for ARPR intensity modulated by optical anisotropy and multilayer interference, which can be applied to $\alpha$-MoO$_3$ flakes with thickness of $d_{\rm MoO_3}$ as follows (detailed discussions in Supplementary Section 3):

\begin{equation}
  \begin{aligned}
     I(\theta) \propto &\int_0^{d_{\rm MoO_3}} |\textit{\textbf{e}}_{s}^\dagger \textit{\textbf{F}}_{s}^\dagger(z) \textit{\textbf{R}}^{\rm int} \textit{\textbf{F}}_{l}(z) \textit{\textbf{e}}_{l}|^2 {\rm d}z.
      \label{ARPRint}
  \end{aligned}
\end{equation}

\noindent $\textit{\textbf{R}}^{\rm int}$ refers to the intrinsic Raman tensor linked to phonon-induced polarizability changes, whose tensor elements are $R^{\rm int}_{xx}$=$|a^{\rm int}|{\rm e}^{{\rm i}\phi^{\rm int}_a}$, $R^{\rm int}_{yy}$=$|b^{\rm int}|{\rm e}^{{\rm i}\phi^{\rm int}_b}$, $R^{\rm int}_{zz}$=$|c^{\rm int}|{\rm e}^{{\rm i}\phi^{\rm int}_c}$ for $A_{\rm g}$ mode.  $\textit{\textbf{F}}_{ l(s)}(z)$ is a diagonal matrix defined as follows  (details in Supplementary Section 3):
\begin{equation}
    \begin{aligned}
    &\textit{\textbf{F}}_{l(s)}(z)={
    \left( \begin{array}{ccc}
    F_{{l(s)},x}(z) & 0 & 0\\
    0 & F_{{l(s)},y}(z) & 0\\
    0 & 0 & 1
    \end{array}
    \right)},
    \end{aligned}
    \label{Factor}
\end{equation}
\noindent where $F_{{l(s)},x}(z)$ and $F_{{l(s)},y}(z)$ describe cavity-modulated field enhancement factors for incident-laser ($l$) and Raman-scattered ($s$) light along the $x$ and $y$ axes, respectively, and are calculated using the TMM\cite{Xie-2025-AM}. $\textit{\textbf{e}}_{s}^\dagger$ and $\textit{\textbf{F}}_{s}^\dagger(z)$ are Hermitian conjugate of $\textit{\textbf{e}}_{s}$ and $\textit{\textbf{F}}_{s}(z)$, respectively. For polarization aligned along the $a$ and $b$ axes of $\alpha$-MoO$_3$, Equation (\ref{ARPRint}) can be simplified as\cite{Xie-2025-AM},

\begin{equation}
  \begin{aligned}
    I_{a(b)} &\propto |a(b)^{\rm int}|^2 \int_0^{d_{\rm MoO_3}} |F_{{s},x}^\dagger(z) F_{{l},x}(z)|^2 {\rm d}z\\
    &\equiv |a(b)^{\rm int}|^2 F_{a(b)},
  \label{Iab}
  \end{aligned}
\end{equation}

\begin{equation}
  \begin{aligned}
    F_{a(b)} = \frac{\int_0^{d_{\rm MoO_3}}\left|t_{12,{l}}t_{21,{s}}A_{l,a(b)}(z)A_{s,a(b)}(z)\right|^2 {\rm d}z}{\left|R_{l,a(b)}R_{s,a(b)}\right|^2},
    \label{EnhanceFact}
  \end{aligned}
\end{equation}

\begin{equation}
  \begin{aligned}
    A_{l,a(b)}(z)&\equiv {\rm e}^{{\rm i}\delta_{{l},a(b)}(z)}+r_{234,{l}}{\rm e}^{{\rm i}\delta_{{l},a(b)}(2d_{\rm MoO_3}-z)},\\
    A_{s,a(b)}(z)&\equiv {\rm e}^{{\rm i}\delta_{{s},a(b)}(z)}+r_{234,{s}}{\rm e}^{{\rm i}\delta_{{s},a(b)}(2d_{\rm MoO_3}-z)},\\
    B_{l,a(b)}&\equiv 1+r_{12,{l}}r_{234,{l}}{\rm e}^{{\rm i}\delta_{{l},a(b)}(2d_{\rm MoO_3})},\\
    B_{s,a(b)}&\equiv 1+r_{12,{s}}r_{234,{s}}{\rm e}^{{\rm i}\delta_{{s},a(b)}(2d_{\rm MoO_3})},
    \label{AR}
  \end{aligned}
\end{equation}

\noindent where $F_{a(b)}$ denotes the Raman enhancement factor along the $a$($b$)-axis; $A_{l(s),a(b)}(z)$ is the electric field component at depth $z$ of the incident-laser (Raman-scattered) light, with a corresponding phase term $\delta_{l(s),a(b)}(z) = 2\pi n_{a(b)}z/\lambda_{l(s)}$; and $B_{l(s),a(b)}$ are the Fabry-P${\rm \acute{e}}$rot interference factors. The coefficients $r_{uv} = (\tilde{n}_u - \tilde{n}_v)/(\tilde{n}_u + \tilde{n}_v)$ and $t_{uv} = 2\tilde{n}_u/(\tilde{n}_u + \tilde{n}_v)$ describe interface reflection and transmission between media $u$ and $v$ ($u,v$ = 1-4 for air, $\alpha$-MoO$_3$, SiO$_2$, Si), respectively, while $r_{234}$ is the effective reflection coefficient at the $\alpha$-MoO$_3$/SiO$_2$/Si interface.

In most common cases, the minimal energy of optical phonons results in an approximation $\lambda_{l}\approx\lambda_{s}$\cite{Xie-2025-AM,Lin-PRL-2025} to simplify Equation (\ref{EnhanceFact}). Based on this assumption, the intrinsic amplitude ratios $|a^{\rm int}|/|b^{\rm int}|$ for the $A_{\rm g}^1$ and $A_{\rm g}^8$ modes at $\lambda_{l}$=$\lambda_{s}$=~532 nm (Supplementary Table S3) were obtained by fitting the ARPR intensities of an $\alpha$-MoO$_3$ flake with specific thickness $d_{\rm MoO_3}$, e.g., $d_{\rm MoO_3}=$ 452 nm in Supplementary Figure S3. With the determined $n_a$, $n_b$ and $|a^{\rm int}|/|b^{\rm int}|$, the $I_a/I_b$ ratios for both modes were then calculated as functions of $d_{\rm MoO_3}$, as indicated by the gray dashed lines in Figure \ref{Fig2}c,d. The oscillatory $I_a/I_b$ ratios with $d_{\rm MoO_3}$ further confirms the cavity effect on polarized Raman intensity driven by birefringence. The calculated $I_a(A_{\rm g}^1)/I_b(A_{\rm g}^1)$ values show good agreement with experimental data up to $d_{\rm MoO_3}$=2000~nm (Figure \ref{Fig2}c); however, significant deviations between the simulated (gray line) and experimental ones are observed for the $A_{\rm g}^8$ mode (Figure \ref{Fig2}d).

Based on the deviations observed in Figure \ref{Fig2}d, we conclude that the wavelength difference between $\lambda_{l}$ and $\lambda_{s}$ must be incorporated when modeling the birefringence-driven optical cavity of $\alpha$-MoO$_3$, particularly for large-wavenumber Raman modes.  By accounting for $\lambda_{l} \neq \lambda_{s}$, we derived the intrinsic Raman tensor element ratios ($|a^{\rm int}|/|b^{\rm int}|$ and $\phi^{\rm int}_{ab}$=$\phi^{\rm int}_{a}-\phi^{\rm int}_{b}$ is the phase difference between $R^{\rm int}_{xx}$ and $R^{\rm int}_{yy}$) for both the $A{\rm g}^1$ and $A_{\rm g}^8$ modes (Supplementary Table S3), enabling quantitative prediction of their $I_a/I_b$ ratios (red and pink lines in Figure \ref{Fig2}c,\ref{Fig2}d, respectively). For the $A_{\rm g}^1$ mode, the calculated $I_a/I_b$ agrees well with the $\lambda_{s} \approx \lambda_{l}$ approximation up to 3000 nm due to its small phonon energy, with only minor deviations beyond this range. In contrast, the $A_{\rm g}^8$ mode shows significant deviation from the approximate result while maintaining excellent agreement with experimental data up to $d_{\rm MoO_3} \sim$ 1000 nm (Figure \ref{Fig2}d). Notably, $I_a(A_{\rm g}^1)/I_b(A_{\rm g}^1)$ exhibits substantial variation (0.07-5) between 900-1000 nm (Figure \ref{Fig2}c), while $I_a(A_{\rm g}^8)/I_b(A_{\rm g}^8)$ remains nearly uniform, consistent with the weak thickness dependence observed experimentally in Figure \ref{Fig2}d. This behavior further confirms the mode-sensitive cavity enhancement of the Raman intensity and the complex modulation mechanism arising from birefringence in such cavity-modulated polarized Raman responses.

The calculated $d_{\rm MoO_3}$-dependent $I_a/I_b$ for both modes exhibit oscillatory ratios with short-period variations superimposed on long-scale modulations. However, their characteristic long-scale modulation periods differ significantly, approximately 2000 nm for the $A_{\rm g}^1$ mode ($\lambda_{s} = 535$ nm) versus about 1000 nm for the $A_{\rm g}^8$ mode ($\lambda_{s} = 563$ nm). This factor-of-two difference reveals a fundamental distinction in their underlying physical origins: the long-scale modulation for the $A_{\rm g}^1$ mode arises directly from birefringence, whereas for the $A_{\rm g}^8$ mode, it necessitates an additional mechanism beyond birefringence, as detailed in the following discussion.

In principle, due to different  wavelengths, the incident-laser and Raman-scattered light should not interfere directly, resulting in separate cavity resonances for each. We calculated the $F_a$ and $F_b$ via Equation (\ref{EnhanceFact}) for the $A_{\rm g}^1$ and $A_{\rm g}^8$ modes, as shown in Figure~\ref{Fig2}f. The oscillatory behavior is governed by the Fabry-P${\rm \acute{e}}$rot interference factors in the denominator of this equation, where the terms $R_{l,a(b)}$ and $R_{s,a(b)}$ determine the characteristic thicknesses for the cavity resonances of incident-laser and Raman-scattered photons, respectively.

For the $A_{\rm g}^1$ mode, the minimal difference between $\lambda_{l}$ and $\lambda_{s}$ results in nearly synchronous cavity resonances. Consequently, both $F_a$ and $F_b$ exhibit intense, narrow resonance peaks across the entire $d_{\rm MoO_3}$ range, as shown in Figure \ref{Fig2}f(i). The short-period oscillations in $F_a$ and $F_b$ correspond to characteristic thicknesses of $d_{l,a}=108$ nm and $d_{l,b}=114$ nm, consistent with the relation $d_{l,a(b)} = \lambda_{l} / (2n_{a(b)})$. The accumulated phase retardation between the $a$- and $b$-axis components is given by $\delta_{a,{l}}(2d_{\rm MoO_3})-\delta_{b,{l}}(2d_{\rm MoO_3})=4\pi\Delta n d_{\rm MoO_3}/\lambda_{l}\approx \pi$, which finally generates high $F_a/F_b$ contrast and consequently high experimental $I_a/I_b$ contrast at $d_{\rm MoO_3}\sim$1000 nm, enabled by $\lambda_{l}\approx\lambda_{s}$ for the $A_{\rm g}^1$ mode. This high-contrast interference condition repeats with a long-scale period of approximately 2000 nm, in excellent agreement with the rigorous TMM results in Figure \ref{Fig2}f, confirming the birefringence-driven nature of the anisotropic optical cavity for $A_{\rm g}^1$ modes in $\alpha$-MoO$_3$ flakes.

For the $A_{\rm g}^8$ mode, the substantial phonon energy induces a Stokes wavelength shift of 31 nm. This shift desynchronizes the cavity resonance conditions between the incident laser ($d_{l,a(b)} = \lambda_{l}/(2n_{a(b)})$) and the Raman-scattered light ($d_{s,a(b)} = \lambda_{s}/(2n_{a(b)})$). The resulting dephasing between $R_{l,a(b)}$ and $R_{s,a(b)}$ in the denominator of Equation (\ref{EnhanceFact}) generates alternating enhancement and suppression of the Raman intensity. At specific thickness intervals (700-1000 nm and 2750-3400 nm), the spatial overlap of enhancement factors between excitation and Raman scattering  becomes minimal, suppressing cavity enhancement and causing resonance peaks to vanish, as shown in Figure \ref{Fig2}g. These "off-cavity" regions account for the reduced characteristic long-scale modulation period observed for the $A_{\rm g}^8$ mode compared to the $A_{\rm g}^1$ mode. The detailed behavior of the $A_{\rm g}^8$ mode thus reveals subtleties that reflect richer underlying physics, governed by the interplay of dual-wavelength cavity resonances.

As discussed above, Figure \ref{Fig2}d-f clearly reveals the mechanism of birefringence-driven optical cavity modulation on polarized Raman intensity. The ultralow absorption of $\alpha$-MoO$_3$ enables multiple long-scale modulations, distinguishing it from highly dissipative ALMs that typically exhibit only one such modulation period\cite{Xie-2025-AM}. We conclude that the birefringence of ALMs primarily governs the quasi-periodic variation of $I_a/I_b$ with $d_{\rm MoO_3}$, with the precise thickness for maximal anisotropy determined by the cavity resonance conditions at both $\lambda_{l}$ and $\lambda_{s}$. A substantial Raman shift can induce significant desynchronization between cavity resonances for incident and Raman-scattered photons. This desynchronization substantially diminishes cavity enhancement, resulting in a less thickness-sensitive $I_a/I_b$ ratio, as demonstrated by the high-frequency $A_{\rm g}^8$ mode in Figure \ref{Fig2}f. This comprehensive understanding underscores the potential for Raman-mode-dependent optical modulation in birefringent ALM flakes.

\begin{figure}[!t]
    \centering
  {\includegraphics[width=0.5\linewidth]{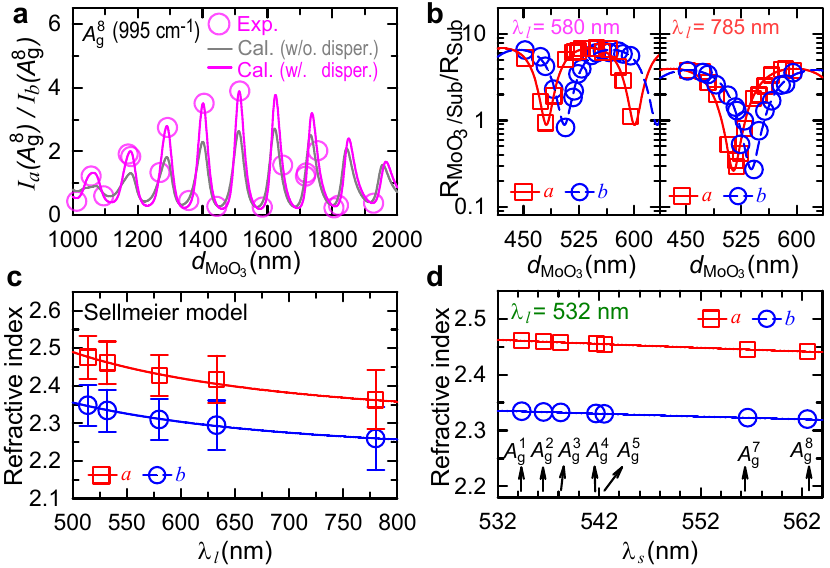}}
  \caption{a) Calculated (curves) and experimental (symbols) $I_a/I_b$ ratios for the $A_{\rm g}^8$ mode versus $d_{\rm MoO_3}$ in the 1000-2000 nm range. Blue and pink curves represent calculations without (w/o) and with (w/) chromatic dispersion at $\lambda_{s}$, respectively. b) Normalized reflectance ($R_{\rm MoO_3/Sub}$/$R_{\rm Sub}$) versus $d_{\rm MoO_3}$ for $\alpha$-MoO$_3$/90 nm-SiO$_2$/Si along the $a$- and $b$-axes relative to the bare substrate at $\lambda_{l}$ = 580 and 785 nm. Symbols: experimental data; curves: fits. c) Sellmeier model fits (curves) to the experimental $n_a$ and $n_b$ of $\alpha$-MoO$_3$. d) $n_a$ and $n_b$ of $\alpha$-MoO$_3$ at $\lambda{_s}$ for each $A_{\rm g}$ mode under $\lambda_{l}$ = 532 nm excitation, with corresponding birefringence values.}\label{Fig3}
\end{figure}

\subsection{Chromatic dispersion-affected birefringence modulation in ARPR Intensity}
The calculated $I_a/I_b$ for the $A_{\rm g}^8$ mode (blue) captures the birefringence-governed periodicity of resonance peaks yet systematically underestimates resonance amplitudes between 1000-2000 nm, as being detailed in $\textbf{Figure}$ \ref{Fig3}a with gray line. We traced this to Equation (\ref{EnhanceFact}), which incorporates $\lambda_s$ but omits the chromatic dispersion of $\alpha$-MoO$_3$, $i.e.$, the wavelength dependence of $n_a$ and $n_b$ for the scattered light. Since chromatic dispersion governed by Kramers-Kronig relations is fundamental to a crystal's optical response, its absence here likely explains the amplitude mismatch\cite{born-2013-book}. Given that the investigated visible range lies below the electronic bandgap and well above the phonon absorption bands of $\alpha$-MoO$_3$\cite{Balendhran-2013-AM,zheng-2019-sciadv}, the dispersion can be approximated by the Sellmeier model, widely used to characterize dielectric responses of transparent crystals\cite{sellmeier-1871-apc} with the typical first-order form:
\begin{equation}
  \begin{aligned}
    n(\lambda) = \sqrt{1+\frac{A \lambda^2}{\lambda^2 - B}},
    \label{Sellmeier}
  \end{aligned}
\end{equation}
\noindent in which $A$ and $B$ are Sellmeier coefficients, and $\lambda$ is the wavelength in micrometers.

To determine these coefficients, we performed polarized reflectance measurements at $\lambda_l$ = 514 nm, 580 nm, 633 nm, and 785 nm, obtaining corresponding $n_a$ and $n_b$ values. Figure \ref{Fig3}b shows experimental reflectance data (symbols) and model fits (solid lines) for $n_a$ and $n_b$ at 580 nm and 785 nm, with additional wavelengths provided in Supplementary Figure S4a,b. The derived $n_a$ and $n_b$ values are summarized in Supplementary Table S2. The Sellmeier-model fits for multi-wavelength data (Figure \ref{Fig3}c), using axis-specific parameters from Supplementary Table S2, are consistent with prior studies\cite{Andres-2021-APL,Shen-2023-2DM} and align well with our DFT calculations (dashed lines in Supplementary Figure S4c).

Using the experimentally parameterized Sellmeier model, Figure \ref{Fig3}d displays $n_a$ and $n_b$ across $\lambda_{\rm s}$ for the $A_{\rm g}^1$ to $A_{\rm g}^8$ modes of $\alpha$-MoO$_3$ under $\lambda_{l} = 532$ nm excitation. By incorporating chromatic dispersion for Raman-scattered photons into Equation (\ref{ARPRint})-(\ref{EnhanceFact}), the intrinsic tensor elements ($|a^{\rm int}|/|b^{\rm int}|$, $\phi^{\rm int}_{ab}$) for the $A_{\rm g}^8$ mode are refined and listed in Supplementary Table S4. With the refined $|a^{\rm int}|/|b^{\rm int}|$, the recalculated $d_{\rm MoO_3}$-dependent $I_a/I_b$ (pink curve) in Figure \ref{Fig3}a now fully reproduces the experimental data across the 1000-2000 nm range. This unified framework also enables to describe the thickness evolution of $I_a/I_b$ ratios for all observed $A_{\rm g}$ modes (see Supplementary Figure S5).

In the ALM community, the small Raman shifts of most materials have fundamentally limited understanding of chromatic dispersion in Raman scattering, as their subtle effects are challenging to resolve experimentally. As shown for $\alpha$-MoO$_3$ in Figure \ref{Fig3}d, dispersion-induced refractive index changes remain modest ($\leq$3$\%$) even for the $A_{\rm g}^8$ mode. Resolving cavity-modulated Raman intensities under such weak dispersion demands exceptionally thick flakes to accumulate detectable phase shifts, e.g., a noticeable effect of optical dispersion on the Raman anisotropy of the $A_{\rm g}^8$ mode emerges at thicknesses above $\sim$1200 nm at 532 nm excitation. However, in narrow-bandgap ALMs like black phosphorus and ReX$_2$ (X=S, Se)\cite{Qiao-2014-NC,Oliva-2019-npj2DM}, strong optical absorption severely limits the penetration depth, suppressing cavity effects in thick flakes and masking dispersion-related phenomena. In contrast, birefringent $\alpha$-MoO$_3$ provides a unique platform combining large Raman shifts with ultralow visible-range absorption, enabling the investigation of cavity-modulated polarized Raman scattering governed by these otherwise subtle optical effects.

\subsection{ARPR intensity for crystallographic orientation identification}

ARPR spectroscopy is widely used to determine the crystallographic orientation of ALMs\cite{Kim-nanoscale-2015,Choi2020NH,Zou-2021-NanoscaleHoz}; however, it mostly depends on sample thickness and substrates. For $\alpha$-MoO$_3$, the evolution of ARPR intensity profile arises from the interplay between the birefringence-driven anisotropic optical cavity effect and the intrinsic in-plane phonon anisotropy. For the  $A_{\rm g}$ modes in $\alpha$-MoO$_3$, the averaged intrinsic in-plane Raman tensor elements ($|a^{\rm int}|/|b^{\rm int}|$ and $\phi_{ab}^{\rm int}$) at $\lambda_l = 532$ nm can be extracted by fitting the corresponding ARPR intensities of samples with different $d_{\rm MoO_3}$ in Supplementary Figure S3. All the experimentally-determined tensor elements are statistically visualized in $\textbf{Figure}$ \ref{Fig4}a,b, and tabulated in Supplementary Table S4. Due to the attenuated signal, $\textbf{\textit{R}}^{\rm int}$ of the $A_{\rm g}^6$ mode is precluded.

\begin{figure*}[!t]
  \centering{\includegraphics[width=1\linewidth]{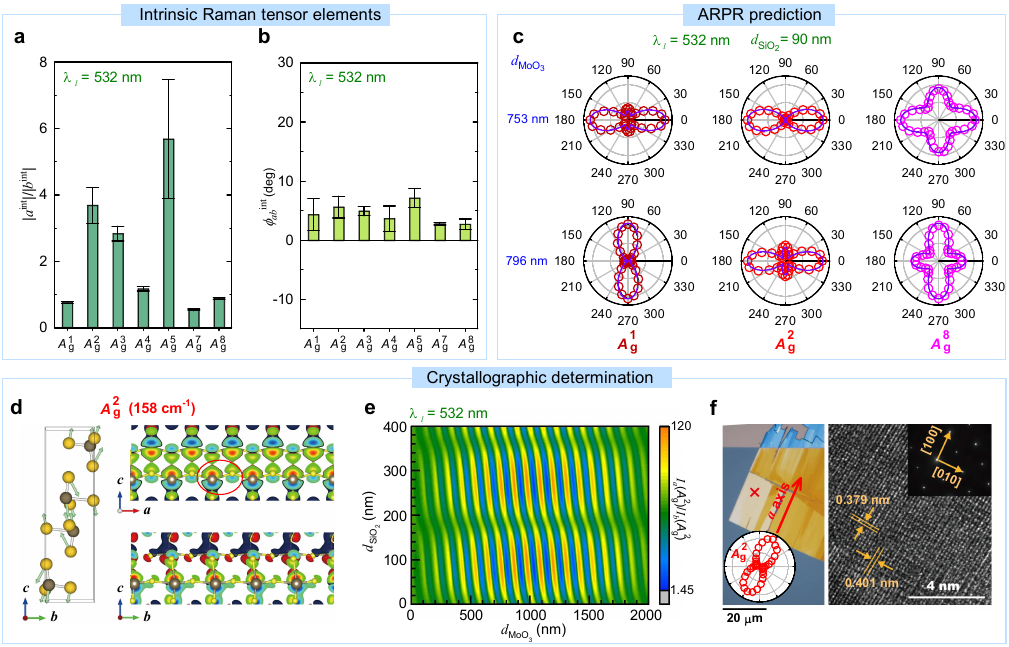}}
  \caption{Mean values of a) $|a^{\rm int}|/|b^{\rm int}|$ and b) $\phi_{ab}^{\rm int}$ for various $\alpha$-MoO$_3$ Raman modes at $\lambda{_l}$ = 532 nm. c) Predicted ARPR responses for $A_{\rm g}^1$, $A_{\rm g}^2$ and $A_{\rm g}^8$ modes in $\alpha$-MoO$_3$/90 nm-SiO$_2$/Si structures with $d_{\rm MoO_3}$ of 753 and 796 nm at $\lambda{_l}$ = 532 nm. d) Contour plot of $I_a/I_b$ for $A_{\rm g}^2$ mode with varied $d_{\rm MoO_3}$ and $d_{\rm SiO_2}$ at $\lambda_l$ = 532 nm. e) Charge density difference of conduction band states before (upper) and after (lower) atomic vibration of the $A_{\rm g}^2$ mode. Isosurfaces: 0.005 e{\AA}$^{-3}$. f) Optical microscopy image of $\alpha$-MoO$_3$ with crystal orientation determined by $A_{\rm g}^2$ ARPR intensity mapping and TEM-SAED.}\label{Fig4}
\end{figure*}

The magnitude ratio $|a^{\rm int}|/|b^{\rm int}|$ quantifies intrinsic in-plane phonon anisotropy for each mode, directly reflecting the anisotropic electron-mediated photon-phonon coupling strength governed by phonon vibrations\cite{Mao-2019-JACS,Zou-2021-NanoscaleHoz}. As shown in Figure \ref{Fig4}a, the $A_{\rm g}^1$, $A_{\rm g}^4$, and $A_{\rm g}^8$ modes exhibit ratios near unity ($|a^{\rm int}|/|b^{\rm int}| \approx 1$), suggesting that the electronic susceptibility changes due to the phonon-induced vibrations exhibit a high degree of isotropy within the $ab$-plane. In contrast, the $A_{\rm g}^2$, $A_{\rm g}^3$, $A_{\rm g}^5$ and $A_{\rm g}^7$ modes display pronounced anisotropy, signifying directionally-dominant electron-photon/electron-phonon coupling along the $a$ axis of $\alpha$-MoO$_3$.

Figure \ref{Fig4}b presents the values of $\phi_{ab}^{\rm int}$ for each $A_{\rm g}$ mode, all of which cluster near 0$^\circ$. This phenomenon stands in great contrast to the previous cases in black phosphorus and T$_d$-WTe$_2$\cite{Xie-2025-AM}. For $\alpha$-MoO$_3$, the below-bandgap excitation at 532 nm excite its electrons into virtual electronic states, resulting in weak electron-photon/phonon interactions\cite{Cardona-1983,Walter-2020-JCTC}, thereby yielding the near-zero $\phi_{ab}^{\rm int}$. In fact, a negligible imaginary part of $\tilde{\epsilon}$ naturally leads to a near-zero value of $\phi_{a(b)}^{\rm int}$\cite{Ribeiro-2015-ACSNano}. With $\phi_{ab}^{\rm int} \approx 0^\circ$, the value of $|a^{\rm int}|/|b^{\rm int}|$ becomes the determining factor for whether the intrinsic phonon anisotropy or the birefringence-induced optical cavity effect governs the $d_{\rm MoO_3}$ evolution of the ARPR intensity profiles.

With $|a^{\rm int}|/|b^{\rm int}|$ (Figure \ref{Fig4}a) and $\phi_{ab}^{\rm int}$ (Figure \ref{Fig4}b), the experimental ARPR intensity profile for each $A_{\rm g}$ mode is accurately described by Equation (\ref{Iab}) (solid lines in Figure \ref{Fig4}c and Supplementary Figure S6a). The ARPR intensity maxima for both $A_{\rm g}^1$ and $A_{\rm g}^8$ modes convert from the $a$-axis (0$^\circ$/180$^\circ$) at $d_{\rm MoO_3}$ = 753 nm to the $b$-axis (90$^\circ$/270$^\circ$) at $d_{\rm MoO_3}$ = 796 nm. This reversal is the direct consequence of their near-unity $|a^{\rm int}|/|b^{\rm int}|$ ratio (Figure \ref{Fig4}a), which enables the Raman enhancement ratio $F_a/F_b$ to dominate over the intrinsic Raman anisotropy for these modes. Furthermore, the $A_{\rm g}^1$ mode exhibits a more pronounced anisotropy change compared to $A_{\rm g}^8$, consistent with its stronger, sharper cavity resonance anisotropy as shown in Figure \ref{Fig2}f.

In stark contrast, the $A_{\rm g}^2$ mode maintains remarkably stable behaviour that its ARPR maximum consistently aligned along the $a$-axis irrespective of $d_{\rm MoO_3}$. Crucially, despite experiencing cavity resonance enhancement comparable to $A_{\rm g}^1$ for their close phonon energies, the large $|a^{\rm int}|/|b^{\rm int}|$ ratio of $A_{\rm g}^2$ mode ensures that the intrinsic phonon anisotropy overrides the birefringence-induced optical cavity effects, endowing $A_{\rm g}^2$ mode with robust ARPR intensity distributions. To elucidate the origin of this large anisotropy in the $A_{\rm g}^2$ mode, we performed DFT calculations for phonon-induced charge density difference in conducted bands, as shown in Figure \ref{Fig4}d. The $A_{\rm g}^2$ mode corresponds to relative translational vibrations of sublayers within each basal plane along the out-of-plane $c$ axis. Despite its out-of-plane vibrational character, the $A_{\rm g}^2$ mode drives substantial in-plane charge density redistribution at bridging oxygen atoms along the $a$ axis (red circles). This directionally-pronounced polarization change results in a far greater modulation of the electronic polarizability derivative along $a$ axis than along $b$ axis, thus a large $|a^{\rm int}|/|b^{\rm int}|$ is expected.

The large $|a^{\rm int}|/|b^{\rm int}|$ of the $A_{\rm g}^2$ mode guarantees its robust $a$-axis alignment of the ARPR intensity maximum (Figure \ref{Fig4}c), enabling its feasibility in assignment of crystallographic orientation. Although the $A_{\rm g}^5$ mode exhibits even higher intrinsic anisotropy, its low scattering intensity (Figure \ref{Fig1}f) makes it less practical for routine measurements. Further experimental and theoretical validation on the 90 nm-SiO$_2$/Si substrates (Supplementary Figure S6b) confirmed persistent $I_a/I_b > 1$ across a wide $d_{\rm MoO_3}$ range. To establish universality across the commonly-used SiO$_2$/Si substrates, we systematically calculated $I_a/I_b$ over $d_{\rm MoO_3}$ and $d_{\rm SiO_2}$ (Figure \ref{Fig4}e), demonstrating substrate-insensitive robustness of $I_a/I_b > 1$. The extreme intrinsic anisotropy ($|a^{\rm int}/b^{\rm int}| \gg 1$) of the $A_{\rm g}^2$ mode provides a reliable indicator for crystallographic orientation determination of $\alpha$-MoO$_3$ flakes on SiO$_2$/Si.

To further confirm the crystallographic orientation determination of MoO$_3$ flakes by $I_a/I_b$ of its $A_{\rm g}^2$ mode, we performed the transmission electron microscopy (TEM) with selected-area electron diffraction (SAED) and ARPR measurements on the same $\alpha$-MoO$_3$ flakes (randomly oriented, Figure \ref{Fig4}f) supported by 30 nm amorphous SiN$_x$ films. The corresponding ARPR intensity profile exhibits a primary lobe that aligns perfectly with the $a$ axis identified by TEM (detailed TEM results in Supplementary Figure S6c; see Supplementary Figure S6d,e for other flakes). This result unequivocally validates the $A_{\rm g}^2$ mode serving as a robust and intrinsic marker for determining in-plane crystallographic orientation.

\section{Conclusion}
In conclusion, this work establishes $\alpha$-MoO$_3$ as a paradigm for birefringence-driven anisotropic optical cavities, leveraging its exceptional in-plane birefringence and ultralow visible-range absorption. We demonstrate that the Fabry-P\'{e}rot cavity formed by $\alpha$-MoO$_3$ flakes on standard substrates enables a unique, mode-selective enhancement of Raman anisotropy, which exhibits a pronounced and complex dependence on flake thickness and Raman shift. A comprehensive quantitative model, incorporating the intrinsic Raman tensor, the wavelength difference between incident and scattered light, and material chromatic dispersion, successfully explains the distinct modulation behaviors observed for different phonon modes, such as the $A_{\rm g}^1$ and $A_{\rm g}^8$ modes. Crucially, the highly anisotropic $A_{\rm g}^2$ mode is identified as a substrate-insensitive intrinsic marker for unambiguous crystallographic orientation determination, validated by cross-correlated TEM and ARPR measurements.

Our findings establish $\alpha$-MoO$_3$ as a model system for probing birefringence-governed light-matter interactions and introduce the optical cavity as a versatile tool for controlling anisotropic Raman responses. The observed interplay between intrinsic phonon anisotropy and birefringence-induced cavity modulation opens new avenues for designing tunable photonic devices and exploring fundamental optical phenomena in low-symmetry van der Waals crystals. This work not only advances the understanding of cavity-enhanced anisotropic spectroscopy but also highlights the potential of low-loss birefringent materials for future on-chip photonic and quantum applications.

\section{Methods}
\subsection{Sample preparation and characterization}
Flakes of $\alpha$-MoO$_3$ were exfoliated from a bulk crystal (HQ graphene) onto polydimethylsiloxane (PDMS) sheets and were subsequently transferred onto SiO$_2$/Si substrates (Si wafers with 90 nm-thick SiO$_2$ on top layer) by the all-dry viscoelastic stamping method\cite{Castellanos-2dm-2014}. The thicknesses of  $\alpha$-MoO$_3$ flakes were measured by the atomic force microscopy (Bruker, Dimension Edge). For TEM measurements, exfoliated $\alpha$-MoO$_3$ flakes were transferred onto Si-based substrates with 30 nm amorphous SiN$_x$ films. TEM and the corresponding SAED images were performed using a spherical aberration corrected TEM (Spectra 300, Thermo Fisher) with 300 kV operating voltage.

\subsection{First-principles calculations}
All first-principles calculations were carried out with VASP\cite{Kresse-1996-CMS} using the projector-augmented-wave formalism and periodic boundary conditions. The exchange-correlation functional for ground-state geometry optimization was PBE within the GGA\cite{Perdew-1996-PRL}. Numerical settings were determined by systematic incrementation of the plane-wave cutoff and $k$-point density, and a setting was declared converged once the change in the primitive-cell total energy between two successive increments was $\leq$ 0.01 eV. $\Gamma$-centered Monkhorst-Pack meshes from 3$\times$3$\times$1 up to 15$\times$15$\times$1 were tested and convergence was achieved at 6$\times$6$\times$1. The electronic threshold was EDIFF = 1$\times$10$^{-8}$ eV; ionic relaxation proceeded until the maximum Hellmann-Feynman force was less than 1$\times$10$^{-8}$ eV {\AA}$^{-1}$. The optimized orthorhombic $\alpha$-MoO$_3$ (Pnma) lattice parameters are $a$ = 3.69 {\AA}, $b$ = 3.92 {\AA} and $c$ = 13.94 {\AA}, in line with prior PBE relaxations and experiment reports for this phase\cite{Sitepu-2009-PD,Wen-2021-JPCC}.

Band structures were computed using a semi-local DFT functional and the HSE06 hybrid functional. To correct DFT's known band-gap underestimation in wide-gap oxides like $\alpha$-MoO$_3$, we employed HSE06 (mixing exact short-range exchange). The resulting 3.25 eV fundamental gap aligns with hybrid-functional benchmarks, providing a reliable baseline for response calculations.

Phonon dispersions were calculated within the harmonic approximation using DFPT (for $\Gamma$-point properties, Born charges $\mathbf{Z}^*$, and $\epsilon_\infty$) and finite-displacement supercells. For polar $\alpha$-MoO$_3$, the non-analytical term correction was applied using $\mathbf{Z}^*$ and $\epsilon_\infty$ to account for LO-TO splitting near $\Gamma$ point. This selectively shifts polar LO branches while leaving non-polar modes robust under our convergence protocol.

Frequency-dependent optical constants were derived from the complex dielectric function $\epsilon(\omega)$. The interband contribution $\epsilon_2(\omega)$ was computed via the Kubo-Greenwood formula using ground-state wavefunctions on dense k-meshes, with $\epsilon_1(\omega)$ obtained by Kramers-Kronig transformation. Where specified, $G_0W_0$ corrections and Bethe-Salpeter equation solutions accounted for quasiparticle and excitonic effects. The complex refractive index followed as $n(\omega)$+${\rm i}\kappa(\omega)$ = $\sqrt{\epsilon(\omega)}$.

\subsection{Raman measurements}
Raman scattering measurements were conducted on a confocal Raman microscope (LabRAM HR Evolution, HORIBA) equipped with a 1800 grooves/mm grating. The 532 nm excitation was from a solid state laser. A 20$\times$ objective with numerical aperture (NA) of 0.25 was used to measure the Raman spectra of the $\alpha$-MoO$_3$ flakes under near-normal incidence on its the basal plane, where the Raman signals were collected in a backscattering geometry under the parallel polarization configuration. The laser power is kept below 2 mW to avoid heating samples. The ARPR spectra were measured by continuously manipulating the polarization angle $\theta$ for the Jones vector of incident laser ($\textbf{\textit{e}}_l$) (and that of Raman scattered light, $\textbf{\textit{e}}_{s}$) relative to the $a$ axis of $\alpha$-MoO$_3$ with a half-wave plate in the common optical path\cite{Xie-2025-AM}. The initial polarization angle ($\theta = 0^\circ$) parallel to $a$ axis of $\alpha$-MoO$_3$ was confirmed for all ARPR measurements.

\subsection{Reflectance measurements}
Polarized reflectance of $\alpha$-MoO$_3$ flakes was measured using the same excitation wavelengths as in the Raman measurements. The incident polarization was purified by a linear polarizer and rotated by a zero-order half-wave plate placed in the common path of the incident and reflected beams. A nonpolarizing beam splitter (Reflectance:Transimittance = 45:55) was utilized to guide the laser onto the measured samples and separate the incident laser from the reflected light. The incident and reflected light intensity were recorded with a power-meter\cite{Xie-2025-AM}.

To determine the refractive indices $n_{a,b}$ along specific crystallographic directions, the laser polarization was aligned with the in-plane axes of $\alpha$-MoO$_3$ to measure the corresponding reflectance from the samples. To correct the light loss from the objective and the offset from the power-meter, the reflectance of $\alpha$-MoO$_3$ on SiO$_2$/Si ($R_{\rm MoO_3/sub}$) was normalized to that of the bare SiO$_2$/Si substrate ($R_{\rm Sub}$) measured at the same focus.
\vspace{1cm}

\raggedright
\medskip
\textbf{Supporting Information} \par 
Supporting Information is available from the Wiley Online Library or from the author.
\vspace{1cm}

\medskip
\textbf{Acknowledgements} \par 
\noindent We acknowledge the support from the National Key Research and Development Program of China (Grant No. 2023YFA1407000), the Strategic Priority Research Program of CAS (Grant No. XDB0460000), National Natural Science Foundation of China (Grant Nos. 12322401, 12127807 and 12393832), Beijing Nova Program (Grant No. 20230484301), Youth Innovation Promotion Association, Chinese Academy of Sciences (No. 2023125) and CAS Project for Young Scientists in Basic Research (YSBR-026).
\vspace{1cm}
\medskip

%

\raggedright
\textbf{Conflict of Interest}\\
\noindent The authors declare no conflict of interest.
\vspace{1cm}

\raggedright
\textbf{Data Availability Statement}\\
\noindent The data that support the findings of this study are available on request from the corresponding author.

\providecommand{\latin}[1]{#1}
\makeatletter
\providecommand{\doi}
  {\begingroup\let\do\@makeother\dospecials
  \catcode`\{=1 \catcode`\}=2 \doi@aux}
\providecommand{\doi@aux}[1]{\endgroup\texttt{#1}}
\makeatother
\providecommand*\mcitethebibliography{\thebibliography}
\csname @ifundefined\endcsname{endmcitethebibliography}
  {\let\endmcitethebibliography\endthebibliography}{}

\end{document}